\documentclass{iopart}
\usepackage[T1]{fontenc}
\usepackage{graphicx}
\usepackage{url,graphics,epsfig}
\usepackage{amssymb}

\usepackage{tikz}

\newcommand*\mycirc[1]{%
  \begin{tikzpicture}[baseline=(C.base)]
    \node[draw,circle,inner sep=1pt](C) {#1};
  \end{tikzpicture}}

\begin{document}

\jl{2}
%
%
%
\def\etal{{\it et al~}}
%
%
%
%
%
%
\setlength{\arraycolsep}{2.5pt}             

\title[K-shell photoionization of Li-like boron ions: Experiment and Theory]{K-shell photoionization of ground-state Li-like boron ions [B$^{2+}$]:
       Experiment and Theory}

\author{A M\"{u}ller$^1$,  S Schippers$^1$, R A Phaneuf$^2$, S W J Scully$^{2,3}$,  A Aguilar$^{2,4}$,  C Cisneros$^5$,
               M F Gharaibeh$^2\footnote[1]{Present address: Department of Physics, Jordan University of Science and Technology,
                                                                           Irbid, 22110, Jordan}$, A S Schlachter$^4$,
              and B M McLaughlin$^{3,6}\footnote[2]{Corresponding author, E-mail: b.mclaughlin@qub.ac.uk}$}

\address{$^1$Institut f\"{u}r Atom- ~und Molek\"{u}lphysik,
                         Justus-Liebig-Universit\"{a}t Giessen, 35392 Giessen, Germany}

\address{$^2$Department of Physics, University of Nevada,
                          Reno, Nevada 89557, USA}

\address{$^3$Centre for Theoretical Atomic, Molecular and Optical Physics (CTAMOP),
                          School of Mathematics and Physics, The David Bates Building, 7 College Park,
                          Queen's University Belfast, Belfast BT7 1NN, UK}

\address{$^4$Advanced Light Source, Lawrence Berkeley National Laboratory, Berkeley, California 94720, USA }

\address{$^5$Centro de Ciencias F\'isicas, Universidad Nacional Aut\'onoma de M\'exico,
                           Apartado Postal 6-96, Cuernavaca 62131, Mexico}

\address{$^6$Institute for Theoretical Atomic and Molecular Physics,
                          Harvard Smithsonian Center for Astrophysics, MS-14,
                          Cambridge, Massachusetts 02138, USA}

%
%
\begin{abstract}
Absolute cross sections for the K-shell photoionization of ground-state
Li-like boron [B$^{2+}$(1s$^2$2s~$^2$S)] ions were measured by
employing the ion-photon merged-beams technique at the
Advanced Light Source synchrotron radiation facility. The energy ranges 197.5--200.5~eV,
201.9--202.1~eV of the [1s(2s\,2p)$^3$P]$^2$P${\rm ^o}$ and
[1s(2s\,2p)$^1$P] $^2$P${\rm ^o}$ resonances, respectively, were
investigated using resolving powers of up to 17\,600.
The energy range of the experiments was extended to about 238.2~eV
yielding energies of the most prominent [1s(2$\ell$\,n$\ell^{\prime}$)]$^2$P$^o$ resonances
with an absolute accuracy of the order of 130 ppm. The natural
linewidths of the [1s(2s\,2p)$^3$P] $^2$P${\rm ^o}$ and [1s(2s\,2p)$^1$P] $^2$P${\rm ^o}$
resonances were measured to be  $4.8 \pm 0.6$~meV and $29.7 \pm 2.5$~meV,
respectively, which compare favourably with  theoretical results of 4.40~meV
and 30.53~meV determined using an intermediate coupling R-matrix method.
\end{abstract}
%
%

\vspace{1.0cm}
\begin{flushleft}
Short title: K-shell photoionization of B$^{2+}$ ions\\
\vspace{1cm} Draft for J. Phys. B: At. Mol. \& Opt. Phys: \today
\end{flushleft}

\maketitle
%
%
%

\section{Introduction}

Satellites such as {\it Chandra} and  {\it XMM-Newton}
are currently providing a wealth of x-ray spectra on many
astronomical objects. There is a serious lack
of adequate atomic data, particularly in the K-shell energy range, needed for
the interpretation of these spectra.
Spectroscopy in the soft x-ray region (5--45~\AA), including
K-shell transitions for singly and multiply charged ionic
forms of atomic elements such as; C, N, O, Ne, S and Si, and the L-shell
transitions of Fe and Ni, provide a valuable probe of the extreme
environments in active galactic nuclei (AGN's),
x~ray binary systems, and cataclysmic variables \cite{McLaughlin2001}.
The goal of the present  experimental and theoretical investigation is to
provide accurate values for photoionization cross sections,
resonance energies, and  linewidths resulting from the
photoabsorption of x~rays near the K-edge of Li-like boron.

Previously, identification of Auger transitions from multiply charged ionic states
of boron has been performed experimentally by using electron
spectroscopy in ion-atom collisions \cite{bruch77,rodbro77,rodbro79},
photon absorption and emission by laser-produced plasmas \cite{kennedy78,Jannitti1984},
electron-impact ionization \cite{Hofmann1990}, beam-foil spectroscopy
and high-resolution spark spectroscopy \cite{Kramida2008}.
Theoretically resonance energies and linewidths for Auger transitions of the Li-like boron ion have been calculated
using a variety of methods such as, perturbation theory in 1/Z \cite{safronova78},
B-splines \cite{Hansen2001} and the saddle-point method (SPM) with
R-matrix or complex co-ordinate rotation methods \cite{chung83,wu91,gd00}.
A critical compilation of atomic transition probabilities for Be and B
covering  all stages of ionization (for allowed and forbidden outer-shell transitions)
has recently been assembled \cite{Fuhr2010}.
To our knowledge this is the first time that measurements of the photoionization cross sections for the
Li-like boron ion have been carried out for the photon energy range in the vicinity of the K-edge.
To complement the high resolution measurements made at the Advanced Light Source (ALS) synchrotron radiation
facility we have carried out photoionization cross section calculations using the R-matrix method.
Recently, high resolution measurements for K-shell photoexcitation of singly, doubly and triply
charged ions of carbon; C$^\mathrm{+}$ \cite{Schlachter2004}, C$^\mathrm{2+}$ \cite{Scully2005} and
C$^\mathrm{3+}$ \cite{Mueller2009} have been carried out within our international collaboration.
A similar approach is used here to perform measurements on the doubly charged boron ion
in the vicinity of the K-edge. Such studies are important in order to provide
accurate results for absolute photoionization cross sections,
resonance energies and natural linewidths. These benchmarked results therefore
update existing literature values \cite{Kasstra1993,rm1979,verner1993,verner1995}
and as such should be used in preference to those that are currently in use
in the various astrophysical modelling codes such as CLOUDY
 \cite{Ferland1998,Ferland2003} and XSTAR \cite{Kallman2001}.

The present study aims to benchmark theoretical values for photoionization cross sections,
resonance energies and lifetimes of autoionizing states of the B$^\mathrm{2+}$ ion in
the vicinity of the K-edge with high-resolution experimental measurements.
This provides confidence in the data that may be used in modelling astrophysical
plasmas; e.g., in the hot (photoionized and collisionally ionized) gas surrounding
$\gamma$ Peg and  $\zeta$ Cas where B\,II and B\,III have been observed in absorption
or for determining boron abundances in early B-type stars in the ORION association \cite{Cunha97,Proffitt99}.

Promotion in Li-like boron (B$^\mathrm{2+}$) ions of a K-shell electron
 to an outer np-valence shell (1s $\rightarrow$ np)  from the ground
state produces states that can autoionize, forming a B$^\mathrm{3+}$ ion and an outgoing free electron.
The strongest excitation process in the interaction of a photon
with the $\rm 1s^22s~^2S$ ground-state of the Li-like boron
ion is the 1s $\rightarrow$ 2p photo-excitation process;
$$
 h\nu + {\rm B^{2+}(1s^22s~^2S)}  \rightarrow  {\rm B^{2+} ~ ([1s(2s\,2p) ~^3P] ~ ^2P^o) }
 $$
 $$
 \downarrow
 $$
 $$
{\rm  B^{3+}~ (1s^2~^1S) + e^-.}
$$
 At higher energies B$^{2+}$([1s(2s\,2p) $^1$P] $^2$P$\rm ^o$) and
 B$^{2+}$([1s\,2$\ell$\,n$\ell^{\prime}]\, ^2$P$\rm ^o$) states (n $\geq$ 3)
are excited which subsequently decay primarily via autoionization processes;
$$
 h\nu + {\rm B^{2+}(1s^22s~^2S)}
 $$
$$
\downarrow
$$
$$
{\rm  B^{2+} [1s(2s\,2p)^1P]~^2P^o, \; {\rm and} \; \;B^{2+} [1s(2\ell\,n\ell^{\prime})^{1,3}P]~^2P^o }
$$
$$
\downarrow
$$
$$
{\rm B^{3+} (1s^2~ ^1S) + e^-.}
$$
The inner-shell autoionization resonances created (by the above processes)
appear in the corresponding photoionization cross sections
(in the energy region near to the K-edge) on top of a
continuous background cross section for direct photoionization of the outer 2s electron. The present investigation provides absolute values (experimental and theoretical) for photoionization cross sections,
resonance energies and linewidths for these states.

The layout of this paper is as follows. Section 2 presents a brief outline of the theoretical work.
Section 3 details the experimental procedure used. Section 4 presents a discussion of the
results obtained from both the experimental and theoretical methods.
Finally in section 5 conclusions are drawn from the present investigation.

\section{Theory}\label{sec:theory}

Theoretical cross-section calculations for the photoionization of doubly charged boron ions are available from the Opacity Project and can be retrieved from the TOPBASE database \cite{Cunto1993}. These cross-section calculations primarily cover the photon energy region corresponding to excitation of valence electrons and have been determined in $LS$-coupling. Theoretical results from the independent particle model exist in the energy region of the K-edge \cite{rm1979,verner1993,verner1995}, but do not account for Auger states that have been observed, e.g.,  in the experimental studies of R{\o}dbro and co-workers \cite{bruch77,rodbro77,rodbro79} or in the present study.  Accurate theoretical estimates for the resonance energies and natural linewidths in the energy region of the K-edge for  the B$^{2+}$  ion have been provided previously by using the saddle-point method in combination with the R-matrix technique \cite{wu91} and  on the basis of the B-spline approach \cite{Hansen2001}. Agreement of that work was found for the  energies of the [1s(2s\,2p)$^{1,3}$P]~$^2$P$\rm ^o$ resonance states with the previous experimental work  of R{\o}dbro and co-workers \cite{rodbro79} and also with earlier theoretical work \cite{safronova78,safronova69,goldsmith74,bhatia78,davis85}.

To benchmark theory for photoionization and obtain suitable agreement with high-resolution experimental
photoionization measurements performed at third-generation synchrotron light source facilities (such as the ALS),
state-of-the-art theoretical methods are required using
highly correlated wavefunctions. Relativistic effects usually need to be included when the
experimental resolution is such that fine-structure effects
can be resolved and radiation damping affects narrow resonances
present in the photoionization cross sections. The features detailed above have been vividly illustrated in numerous experimental and theoretical photoionization studies within the present collaboration. Recent work on K-shell ionization of light ions involved He-like Li$^\mathrm{+}$ \cite{Scully2006,Scully2007}, Li-like C$^{3+}$~\cite{Mueller2009}, Be-like C$^\mathrm{2+}$ \cite{Scully2005}, and  B-like C$^\mathrm{+}$  \cite{Schlachter2004}.

Photoionization cross-section  calculations for  B$^\mathrm{2+}$ ions were
performed both in $LS$ and in intermediate coupling using the semi-relativistic
Breit-Pauli approximation which allows for relativistic effects to be included.
Radiation-damping \cite{damp} effects were also included for completeness within
the confines of the R-matrix approach \cite{rmat,codes}.
An appropriate number of B$^\mathrm{3+}$ states (11 $LS$, 17 $LSJ$ levels) were
included in our intermediate coupling calculations. An n=4 basis set of
B$^\mathrm{3+}$ orbitals was used which was constructed using the
atomic-structure code CIV3 \cite{Hibbert1975} to represent the wavefunctions.
Photoionization cross-section calculations were then performed in intermediate
coupling for the $\rm 1s^22s~^2S_{1/2}$ initial state of the B$^\mathrm{2+}$ ion.
In the calculations the following eleven He-like $LS$ states were retained:
$\rm 1s^2~^1S$,  $\rm 1s\,ns~^{1,3}S$, $\rm 1s\,np~^{1,3}P^{\,\circ}$, and $\rm 1s\,nd~^{1,3}D$, for n$\leq$ 3.
The additional configurations $\rm 1s\,np~^{1,3}P^{\,\circ}$, and $\rm 1s\,nd~^{1,3}D$,
$\rm 1s\,nf~^{1,3}F^{\,\circ}$, (n = 4),
of the B$^\mathrm{3+}$  were used to account for correlation.
This gives rise to 17 $LSJ$ states in the intermediate  close-coupling
expansions for the $J$=1/2 initial scattering symmetry of the Li-like  B$^{2+}$ ion.
For the structure calculations of the B$^\mathrm{3+}$ ion, all physical orbitals were
included up to n=3 in the configuration-interaction wavefunction expansions
used to describe the states.
The Hartree-Fock $\rm 1s$ and $\rm 2s$ tabulated orbitals of Clementi and Roetti
\cite{Clementi1974} were used together with n=3
 orbitals which were determined by energy optimization on the appropriate
spectroscopic state using the atomic structure code CIV3
\cite{Hibbert1975}.  The n=4 correlation (pseudo) orbitals were determined by energy
optimization on the 1s\,2s~$^1$S hole state of the B$^{3+}$ ion in order to
account for core relaxation and additional correlation effects in the
multi-configuration interaction wavefunctions.
All the states of the B$^\mathrm{3+}$ ion were then represented
by using multi-configuration interaction wave functions.  The Breit-Pauli
$R$-matrix approach was used to calculate the energies
of the B${^\mathrm{2+}}(LSJ)$ states and the subsequent photoionization cross sections.
A minor shift ($<$  0.1 \%) of the theoretical energies  to experimental values
was made so that they would be in agreement with
available experimental thresholds \cite{Ralchenko2008}.  Cross sections for photoionization out of the
B$^\mathrm{2+}$ (1s$\rm ^2$2s $\rm ^2$S$_{\rm 1/2}$ ) ground-state ion
were then obtained for total angular momentum scattering symmetries
of $J$ = 1/2 and $J$= 3/2, odd parity, that contribute to the total.

The $R$-Matrix method \cite{rmat,codes,damp} was used to determine
all the photoionization (PI) cross sections $\sigma_{\rm PI}$ (E), for the initial ground state in intermediate-coupling ($LSJ$).
The scattering wavefunctions were generated by
allowing all possible three-electron promotions out of the base $\rm 1s^22s$
configuration of B$^\mathrm{2+}$ into the orbital set employed.
Scattering calculations were performed with forty
continuum functions and a boundary radius of 9.8 Bohr radii.
For the $\rm ^2S_{1/2}$ initial state the outer region electron-ion collision
problem was solved (in the resonance region below and
 between all the thresholds) using a suitably chosen fine
energy mesh of 5$\times$10$^{-8}$ Rydbergs ($\approx$ 0.68 $\mu$eV)
to fully resolve all the extremely fine resonance
structure in the photoionization cross sections.
The multi-channel R-matrix QB technique (applicable to atomic and molecular complexes)
of Berrington and co-workers \cite{keith1996,keith1998,keith1999}
was used to determine the resonance parameters. The resonance width $\Gamma$
was determined from the inverse of the energy derivative of the eigenphase
sum $\delta$ at the position of the resonance energy $E_r$ via
\begin{equation}
\Gamma = 2\left[{\frac{d\delta}{dE}}\right]^{-1}_{E=E_r} = 2 [\delta^{\prime}]^{-1}_{E=E_r} \quad.
\end{equation}
Averaging was performed over final total angular
momentum $J$ values. Finally, in order to compare directly with experiment,
the theoretical cross section was convoluted with a Gaussian
function of appropriate width to simulate the energy resolution of the measurements carried out
at the Advanced Light Source synchrotron radiation facility.

\section{Experiment}\label{sec:exp}

The experiment was performed at the ion-photon-beam (IPB) end-station of the undulator beamline 10.0.1 at the ALS. A detailed description of the experimental setup has been provided by Covington \etal \cite{Covington2002}. For the present study on B$^{2+}$ similar procedures were utilized as in our earlier measurements on the K-shell photoionization cross sections
for carbon ions \cite{Schlachter2004,Scully2005,Mueller2009}. Beams experiments with Li-like ions such as C$^{3+}$ and B$^{2+}$ are attractive since such ions can be provided almost exclusively in their ground state. The metastable $^4$P  states of B$^{2+}$ have lifetimes of less than 300~ns~\cite{Davis1989}, much too short to survive the flight time of approximately 10~$\mu$s from the ion source to the photon-ion interaction region.

The boron ions were generated from BF$_3$ gas inside a compact all permanent-magnet electron cyclotron-resonance (ECR) ion source \cite{Broetz2001}. Collimated B$^{2+}$ ion-beam currents of
typically 30~nA  were extracted by putting the ion source on a
positive potential of +6~kV. After having travelled through a bending
dipole magnet serving to select the desired ratio of charge to
mass, the ion beam was centered onto the counterpropagating photon beam by applying
appropriate voltages to several electrostatic ion beam steering devices. Downstream of  the
interaction region, the ion beam was deflected out of the photon beam direction by a second
dipole magnet that also separated the ionized B$^{3+}$ product ions from the B$^{2+}$ parent
ions. The B$^{3+}$ ions were counted with nearly 100\% efficiency with a single-particle detector,
and the B$^{2+}$ ion current was monitored for normalization. The measured B$^{3+}$
count rate $R$ was only partly due to photoionization events. It also contained B$^{3+}$ ions produced
by electron removal collisions with residual gas molecules and surfaces.
This background was determined by mechanically chopping the photon beam.

Absolute cross sections were measured by normalizing the background-subtracted B$^{3+}$ count rate
to the measured ion current, to the photon flux, which was measured with a calibrated photodiode,
and to the beam overlap. Beam overlap measurements were carried out using two commercial
rotating-wire beam-profile monitors and a movable slit scanner. Absolute cross sections and
resonance strengths for photoionization have thus been obtained with an estimated
uncertainty of 20\%. Detailed descriptions of the procedures for absolute measurements and of the budget of uncertainties have been provided previously by Covington \etal~\cite{Covington2002}. Due to the considerable effort required for carrying out reliable absolute cross section measurements these were only performed at the peak positions of the dominant photoionization resonances.

Energy scan measurements were taken by stepping the photon energy
through a preset range of values. The desired experimental energy spread was preselected by
adjusting monochromator settings of the beamline  accordingly. The scan measurements were
normalized to the absolute data points. The energy scale was calibrated by carrying out photoabsorption measurements
with SF$_6$ and Ar gas for the well known resonance features \cite{Hudson1993,King1977} at
energies 176--182~eV and 242--251~eV, respectively. Calibrated monochromator settings from
these ranges were linearly interpolated to obtain the scaling factors for measured
energies in the present range of interest. Since the parent ions are in motion, a Doppler correction
has to be carried out transforming the nominal laboratory energies to the center-of-mass frame of the ions
before applying the calibration factor. With all this carefully included, we estimate an uncertainty of
at most $\pm 30$~meV for the energy scale of the present measurements.
Since the precision of peak position determinations is of the order of better than 1~meV,
the possible calibration error almost exclusively determines the absolute uncertainties of the resonance energies.

%
%
%

\begin{figure}
\begin{center}
\includegraphics[width=\textwidth]{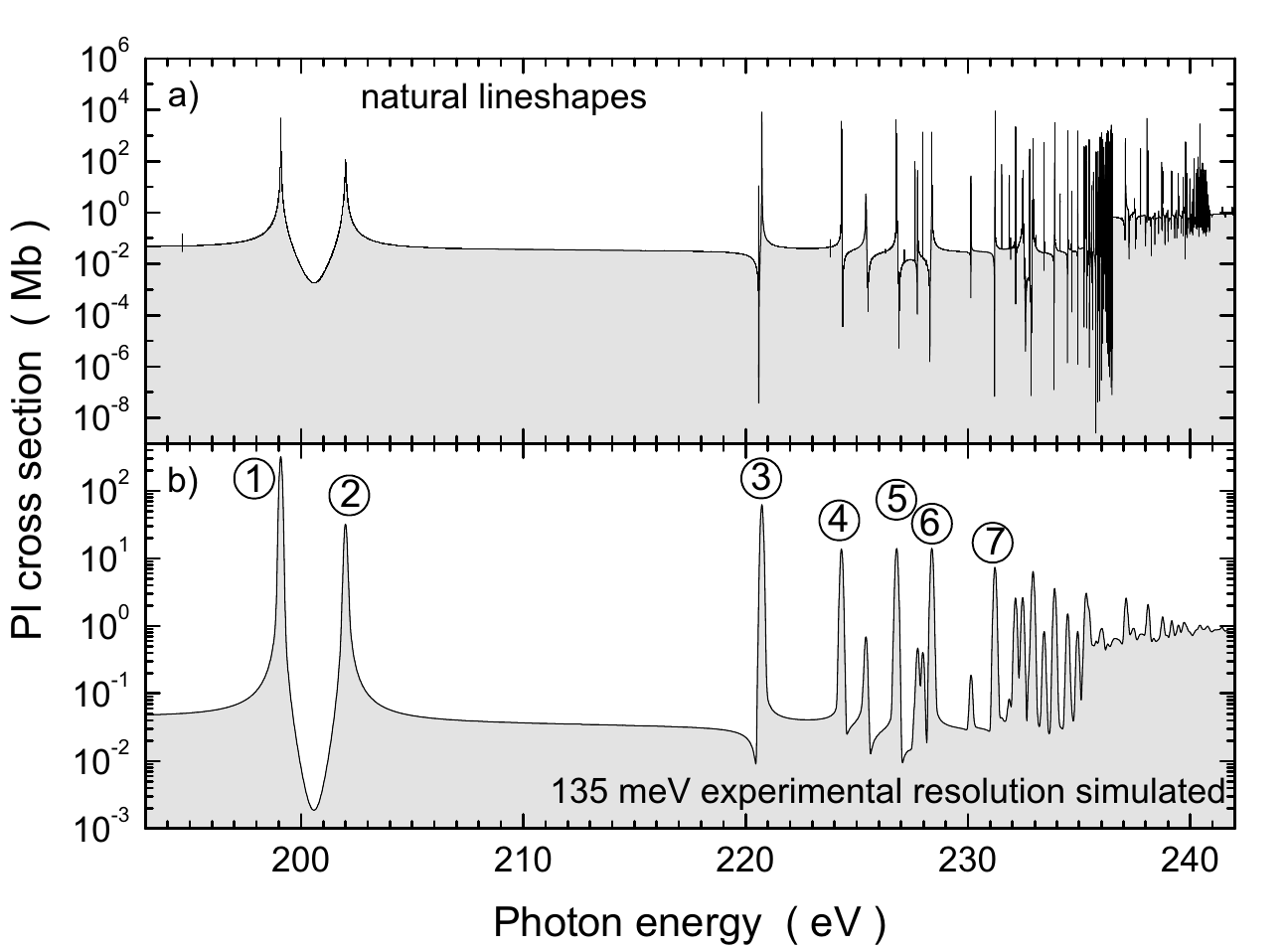}
\caption{\label{fig:Fig1}  Theoretical cross sections for K-shell photoionization (PI) of Li-like B$^\mathrm{2+}$
                                            from the 17-state intermediate-coupling R-matrix calculations;
                                            a)  results showing the natural line shapes and interference of direct and resonant photoionization; b) the same
                                            calculations convoluted with a Gaussian distribution function of 135~meV full width
                                            at half maximum (FWHM). The K-edge at 236.496~eV is evident on the lower panel.
                                            Peaks~\mycirc{1} through \mycirc{7} observed in the experiment are designated respectively as;
                                            [1s(2s\,2p)$^3$P]$\rm ^2P^o$, [1s(2s\,2p)$^1$P]$\rm ^2P^o$, [(1s\,2s $^3$S)3p]$\rm ^2P^o$, [(1s\,2s $^1$S)3p]$\rm ^2P^o$,
                                            [(1s\,2p $^1$P)3s]$\rm ^2P^o$, [(1s\,2p $^1$P)3d]$\rm ^2P^o$ and [1s(3s\,3p) $^1$P]$\rm ^2P^o$. }
\end{center}
\end{figure}

\begin{figure}
\begin{center}
\includegraphics[width=\textwidth]{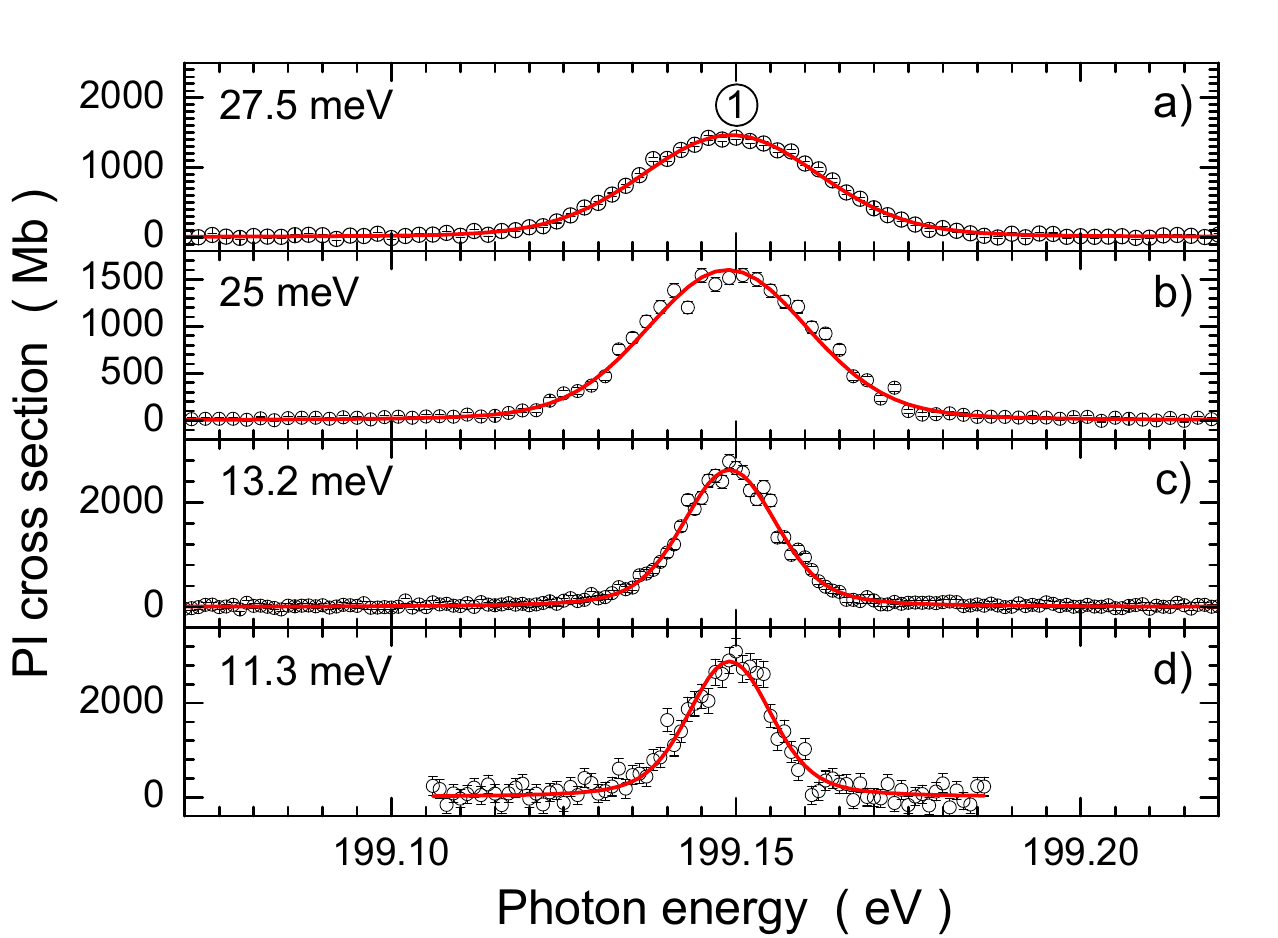}
\caption{\label{fig:Fig2} (Colour online) Series of energy scans taken for peak~\mycirc{1}
                                           in figure~\ref{fig:Fig1}. The scan cross sections were normalized to an absolute
                                           measurement at the peak energy
                                           so that all data sets are on an absolute scale.
                                           The peak is associated with the  [1s\,(2s\,2p)$\rm ^3$P]$\rm ^2$P${\rm ^o}$ resonance
                                           in B$^{2+}$ ions. The experimental energy spreads given for each individual measurement were determined by Voigt fits (red solid lines) to the experimental data.
                                           }
\end{center}
\end{figure}

\begin{figure}
\begin{center}
\includegraphics[width=\textwidth]{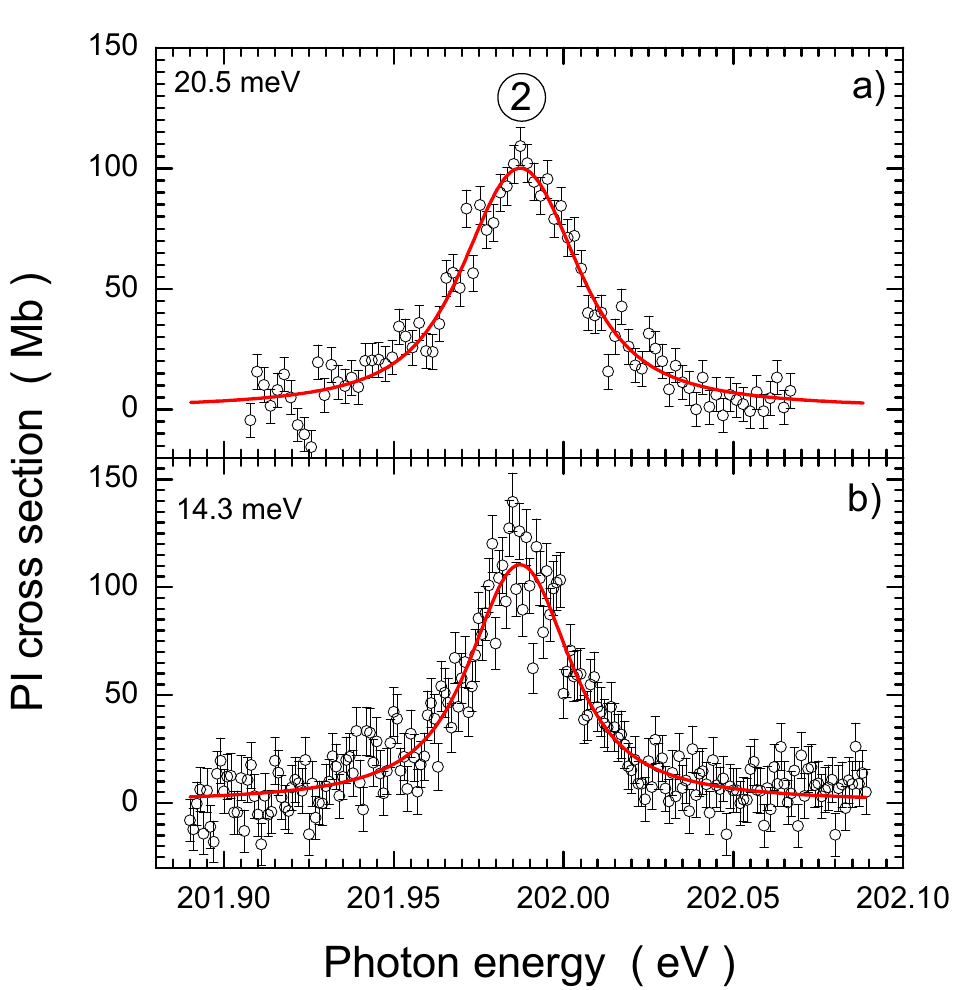}
\caption{\label{fig:Fig3} (Colour online) Energy scans taken for peak~\mycirc{2} in figure~\ref{fig:Fig1}.
                                           Like the data in figure~\ref{fig:Fig2} the cross sections are on an absolute scale with statistical error bars provided. The peak is associated with the  [1s\,(2s\,2p)$\rm ^1$P]$\rm ^2$P${\rm ^o}$ resonance in B$^{2+}$ ions.
                                            The experimental energy spreads  were determined
                                            by Voigt fits (red solid lines) to the experimental data.}
\end{center}
\end{figure}

\begin{figure}
\begin{center}
\includegraphics[width=\textwidth]{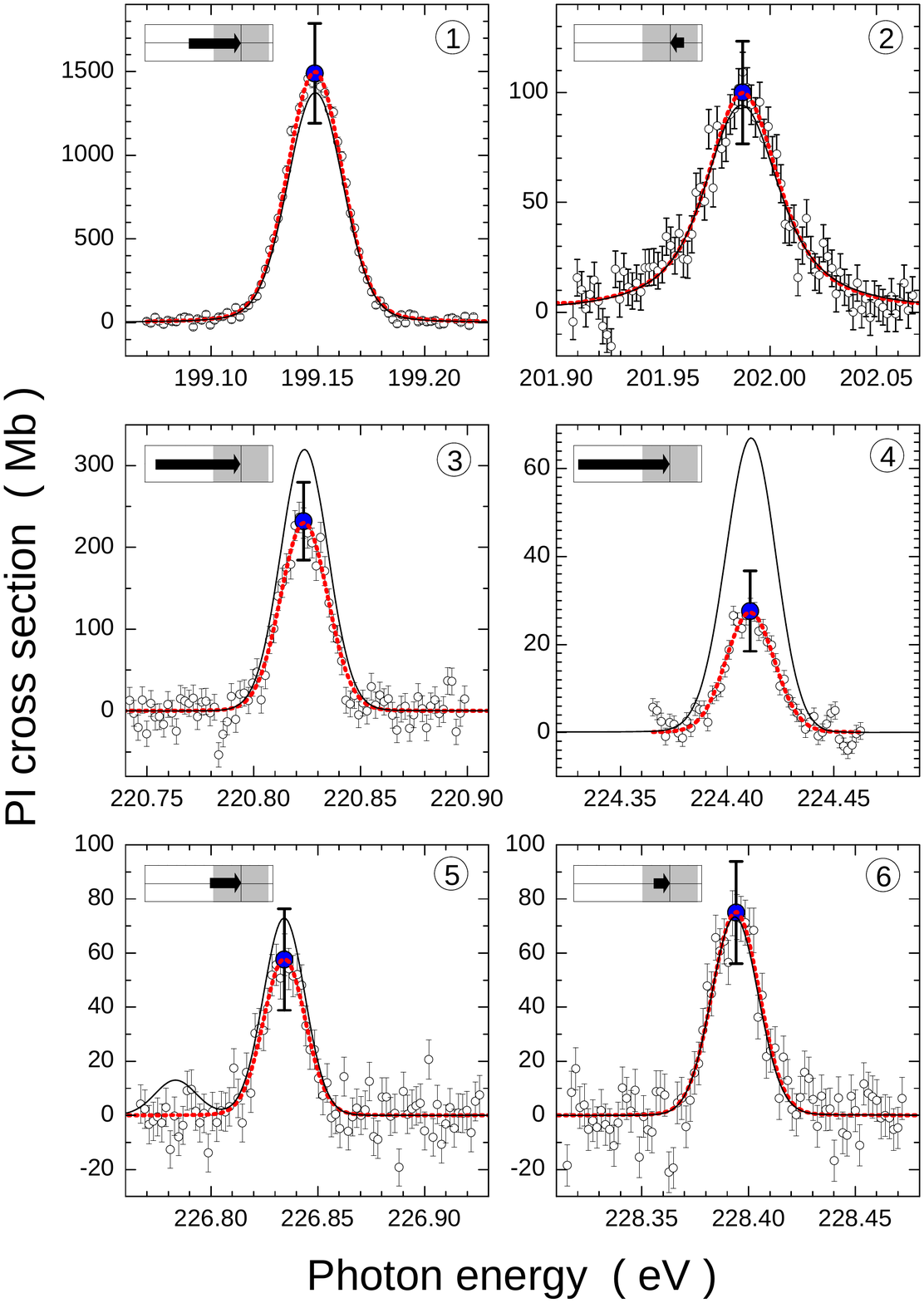}
\caption{\label{fig:Fig4} (Colour online) Energy scan measurements taken for peaks~\mycirc{1}
                                          through \mycirc{6} in figure~\ref{fig:Fig1}. The scan cross sections are shown with statistical uncertainties. The data  were normalized to absolute
                                          measurements at the peak energy (shown as blue shaded circles with fat error bars).
                                          Voigt fits to the experimental data are represented by the (red) dotted lines. The R-matrix results were convoluted with Gaussians of appropriate widths for each individual peak. The theory curves (solid black lines) were then shifted to the measured peak positions. The shifts necessary to match theory with experiment are visualized by the lengths of the arrows inside the boxes in the upper left corner of each panel.                                       Each box represents an energy range from -105~meV to +35~meV around the resonance energy. The gray shaded area represents an experimental uncertainty range of $\pm 30$~meV. An arrow from left to right indicates that the calculated resonance had to be shifted to higher energies.}
\end{center}
\end{figure}

\begin{figure}
\includegraphics[width=\textwidth]{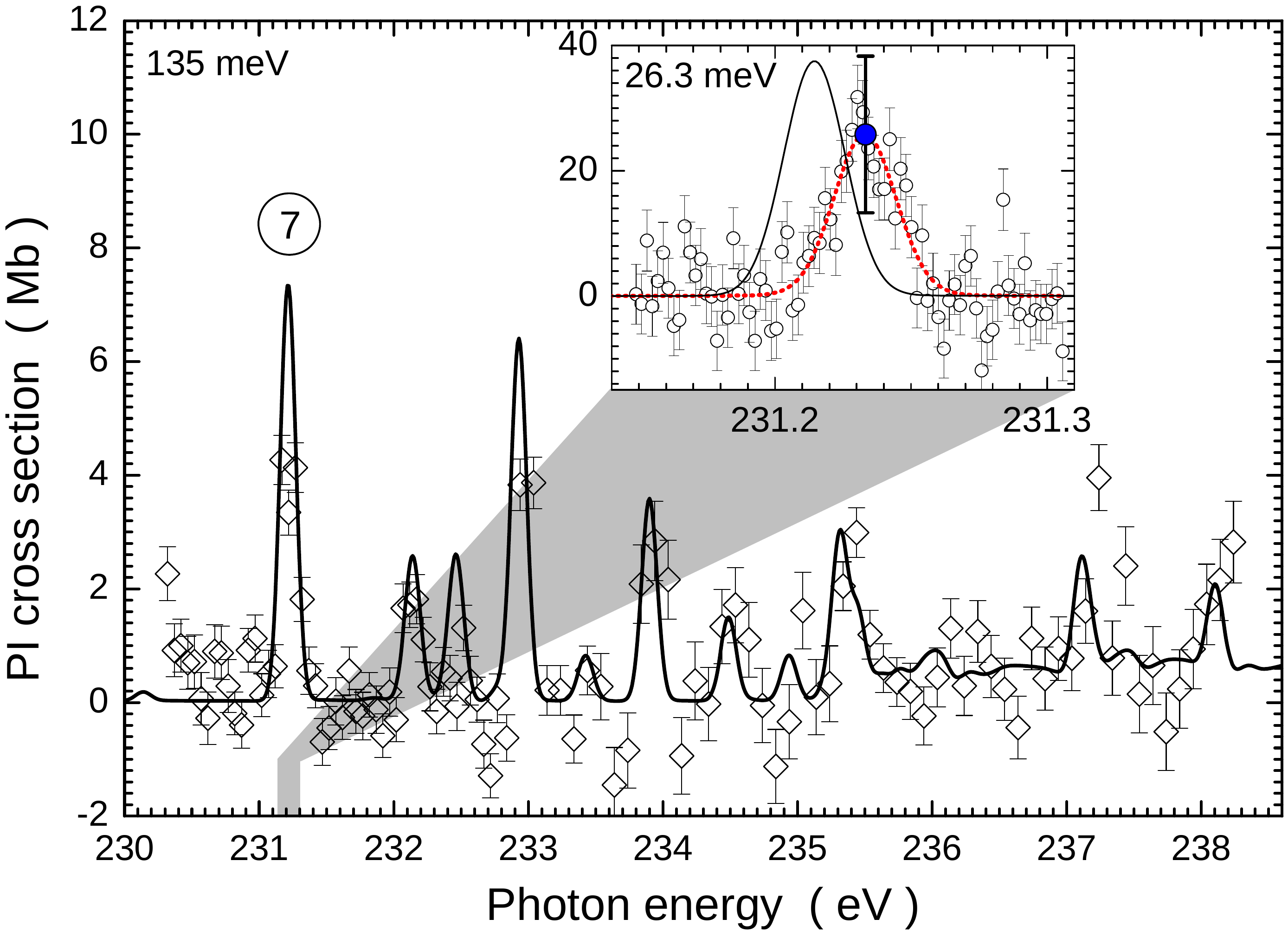}
\caption{\label{fig:Fig5} (Colour online) Energy scan measurement taken in the energy range from
                                           230.3--238.3~eV at an experimental energy spread of 135~meV.
                                           The open diamonds with their statistical error bars were obtained by averaging over 5 adjacent measurements. The scan cross sections were normalized  to the absolute measurement at the centre of peak~\mycirc{7} (shown as blue shaded circle with fat error bars in the inset) so that the data are on an absolute scale.
                                           Peak~\mycirc{7} was separately scanned at a resolution of 26.3~meV (see inset) determined
                                           by a Voigt fit (red dotted line) to the experimental data. The R-matrix results from
                                           the lower panel of figure~\ref{fig:Fig1} are displayed as a solid black line.
                                           The inset also shows the R-matrix results as a non-shifted solid black line
                                           that was obtained by convolution with a Gaussian of 26.3~meV FWHM.
                                           The energy range of the inset is 170~meV, i.e., the same as in all panels
                                           of figure~\ref{fig:Fig4}. The shading indicates the origin of the inset panel
                                           on the energy axis of the present graph.}
\end{figure}

\begin{figure}
\includegraphics[width=\textwidth]{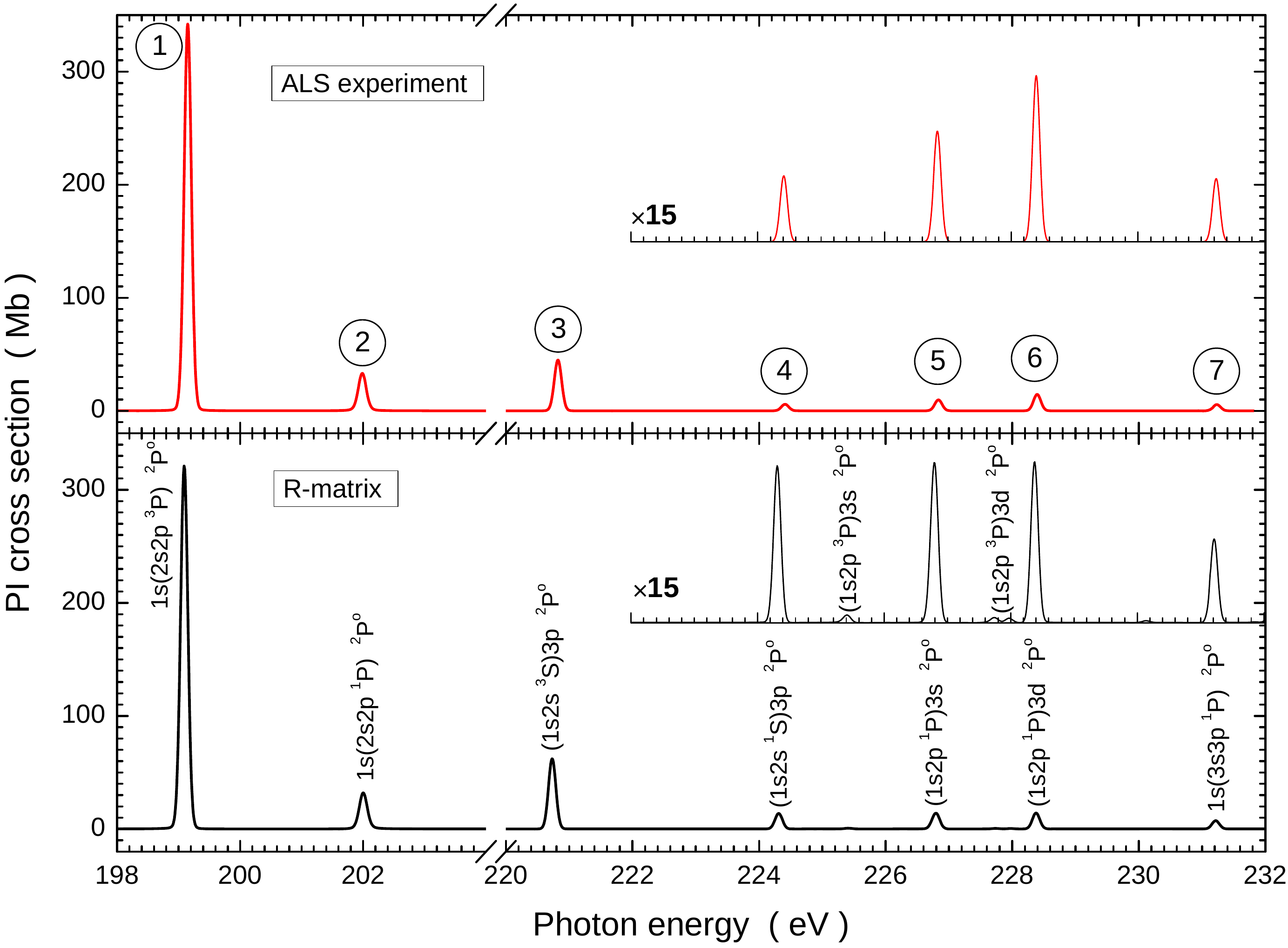}
\caption{\label{fig:Fig6} (Colour online) Comparison of theory and experiment in an overview.
                                           The lower panel shows the theory data taken from the
                                           lower panel of figure~\ref{fig:Fig1} now displayed on a linear scale.
                                           The suggested peak assignments are provided. The cross sections
                                           in the energy range above 222~eV were multiplied by a factor of 15
                                           and displayed with an offset. The red line in the upper panel is based
                                           on fits to the experimental results of this study: the Voigt profiles of resonances
                                           no. 1 through 7 were simulated for an experimental energy spread of 135~meV
                                           using the information obtained from the
                                           Voigt fits to the experimental data displayed in figures~$\rm 2 - 5$.}
\end{figure}

\section{Results and Discussion}\label{sec:res}
Figure 1 shows the theoretical photoionization cross section spectra for the B$^{2+}$ ion obtained from our intermediate-coupling R-matrix results in the photon energy range 193--242~eV. In figure 1 a) the theoretical results are displayed in their natural form. For the graph in figure 1 b) the theoretical results were  convoluted with a Gaussian distribution function of 135~meV full width at half maximum. The resonance peaks labelled~\mycirc{1} through ~\mycirc{7} shown in the convoluted  theoretical photoionization spectrum of figure 1 b) have been observed in the photoionization experimental
data taken at the ALS. On the logarithmic scale of this figure, one can clearly observe interference features resulting from the interaction of direct $\rm 2s$ photoionization and K-shell excitation resonances. The resulting changes at the level of the $\rm 2s$ ionization continuum are so small, however, that they are not accessible to the present experiment.

The experimental  and theoretical results for K-shell photoionization  cross sections of the B$^{2+}$ ion
are shown in figures 2--5. The solid black  lines in figures 4 and 5 are the theoretical results from the R-matrix calculations including radiation damping and convoluted with a Gaussian of the appropriate experimental resolution.
The results  from the ALS experiment in figures 2 -- 5 are indicated by the open circles (diamonds instead of open circles in figure~\ref{fig:Fig5} to indicate the greatly different experimental energy spread in that particular measurement) and the
solid red lines (dotted red lines in figures 4 and 5) represent Voigt fits to the data. From the Voigt fits to the experimental data experimental energy spreads were determined. These energy spreads are given in figures~2 and 3 in each panel. For figure 4 they are 27.5~meV for peak~\mycirc{1}, 20.5~meV for peak~\mycirc{2}, 25.5~meV for peak~\mycirc{3}, 26~meV for peak~\mycirc{4}, 22.6~meV for peak~\mycirc{5}, and 25.4~meV for peak~\mycirc{6}. In figure 5 a scan measurement with 135 meV energy spread is shown. The inset shows peak~\mycirc{7} again but this time measured with an experimental energy spread of 26.3 meV.

Figures 2 and 3 show the ALS experimental data for the B$^{2+}$ ion in the energy region of the first two
resonance peaks (the [1s\,(2s\,2p)$^{1,3}$P]$^2$P$\rm ^o$ resonances) taken during the course of the measurements with varying degrees of resolution. By fitting Voigt profiles to the different sets of peak measurements the experimental energy spreads, the natural resonance widths and resonance energies can be determined. Each single scan of a resonance can contribute to the accuracy of the resonance parameters obtained. The lowest-energy-spread experiments are especially valuable for the determination of the natural line width. The measurement with the smallest energy spread may, however, suffer from statistics and thus a higher--energy-spread experiment with better statistics may add significantly to the accurate determination of the natural resonance width and resonance energy. Thus for peak~\mycirc{1} we obtained widths $\Gamma  = 4.5 \pm 1.7$~meV from the experimental data shown in figure~2 panel d) at the highest resolution and $\Gamma  = 5.2 \pm 0.5$~meV from the experimental data shown in figure~2 panel c). The data from figure~2 panels a) and b) where inconclusive with respect to the line width when fitted individually. Similarly, for peak~\mycirc{2} we obtained widths $\Gamma  = 29.5 \pm 3.3$~meV from the experimental data shown in figure~3 panel b) and $\Gamma  = 30.2 \pm 3.9$~meV from the experimental data shown in figure~3 panel a).

Clearly, the result obtained from several independent measurements can be expected to be of better quality than the results of each single measurement. Therefore, by combining all data sets obtained for a specific resonance in a multiple-peak fit one can improve the accuracy of the final result. In such a least squares multiple-peak fit all measurements are considered in a single fit procedure. The resonance energy $E_{\rm ph}^{(res)}$ and the natural width $\Gamma $ of the resonance investigated in several individual measurements are varied in order to minimize the sum of all squared deviations in all measurements at a time. Of course, they have to be  kept the same for all the different individual peak measurements on one specific resonance. Fitting the 4 measurements of peak~\mycirc{1} shown in figure~2 in one least-squares fit session provides $\Gamma  = 4.8 \pm 0.6$~meV. Similarly, a combined fit of the 2 measurements of peak~\mycirc{2} in figure~3 yields $\Gamma  = 29.7 \pm 2.5$~meV. While the fit results obtained for the first two peaks are of very satisfying quality, Fano parameters describing the interference of direct and resonant photoionization channels  were not accessible to the present experiment. The results that could be obtained from these nonlinear least-squares fits to the measured data illustrated in figures~2 and 3 are presented in table \ref{tab:fit1}. Apart from the resonance energies $E_{\rm ph}^{(res)}$ and the Lorentzian widths $\Gamma $, experimental resonance strengths $\overline{\sigma}^{\rm PI}$ were also obtained which result from integration of the cross section $\sigma_{\rm PI} ({\rm E}; L_i S_i \rightarrow L_j S_j )$ for a specific resonance transition over all energies
\begin{equation}
\overline{\sigma}^{\rm PI} = \int_{E}{}  \sigma_{\rm PI} ({\rm E}; L_i S_i \rightarrow L_j S_j )d{\rm E}\quad.
\end{equation}

%
%
%

\begin{table}
\caption{\label{tab:fit1} Comparison of photoionization resonance energies $E_{\rm ph}^{\rm (res)}$ (in eV),
         natural linewidths $\Gamma$ (meV) and resonance strengths $\overline{\sigma}^{\rm PI}$ (in Mb~eV)
         for B$^{2+}$ ions from our present  ALS experimental work and
         R-matrix theoretical (11-state $LS$-coupling and 17-state intermediate coupling)
         calculations with previous studies for peaks~\mycirc{1} and~\mycirc{2}. For the conversion of the wavelength scale of
         some of the previous work to the present energy scale $hc = 1239.841\,875$~eV~nm has been used. For the conversion of B$^{2+}$ Auger energies obtained by Chung and Bruch~\cite{chung83} to the B$^{2+}$ excitation energies in the present table the ionization potential of 37.9306~eV for B$^{2+}$ ions as tabulated by Ralchenko \etal~\cite{Ralchenko2008}  was added.}
\begin{indented}
 \lineup
 \item[]\begin{tabular}{ccr@{\,}c@{\,}llcl}
\br
 Resonance    & & \multicolumn{3}{c}{ALS/Others}   & \multicolumn{1}{c}{R-matrix} & \multicolumn{2}{c}{SPM/B-spline}\\
 (Label)      & & \multicolumn{3}{c}{(Experiment)} & \multicolumn{1}{c}{(Theory)} & \multicolumn{2}{c}{(Theory)}\\
 \ns
 \mr
 $$[1s(2s\,2p)$^3$P]$^2$P$\rm ^o$ & $E_{\rm ph}^{\rm (res)}$
             	& 199.149  &$\pm$  			& \00.030$^{\dagger}$ 	& 199.092$^{a}$  & &199.279$^{c}$ \\
  ~\mycirc{1}     &     	& 199.130  &$\pm$  			& \00.010$^{f}$     		 & 199.081$^{b}$  & &199.101$^{d}$ \\
  &          	& 199.17\0 &$\pm$  			& \00.03$^g$       		&                		    & &199.145$^{e}$ \\
  &          	& 199.17\0 &$\pm$  			& \00.1$^h$        		 &                  	    & &199.181$^{l}$  \\
  &          	& \multicolumn{2}{l}{199.20$^{k}$}  & 			 	&                	             & &              \\
  \\
 & $\Gamma$
             & 4.8\0\0  &$\pm$  & \00.6$^{\dagger}$   	& \0\04.40$^{a}$ 	& 	 &\0\05.27$^{c}$ \\
 &           &          &       &                     				& \0\04.45$^{b}$ 	& 	&\0\04.00$^{d}$ \\
 &           &          &       &                     				&               			 &	 &\0\04.29$^{e}$ \\
 &           &          &       &                     				&               			 &	 &\0\04.05$^{l}$ \\
  \\
 & $\overline{\sigma}^{\rm PI}$
             & 51\;\0\0\0     &$\pm$  & 10$^{\dagger}$      & \045.34$^{a}$  & &\\
 &           &          &       &                     & \045.66$^{b}$  & &\\
 \\
 \\
 $$[1s(2s\,2p)$^1$P]$^2$P$\rm ^o$ & $E_{\rm ph}^{\rm (res)}$
             & 201.987 &$\pm$  &\00.030$^{\dagger}$   & 201.991$^{a}$   	& & 202.227$^{c}$   \\
  ~\mycirc{2}  &        & 201.976 &$\pm$  &\00.014$^{f}$         & 201.941$^{b}$         	 & & 201.987$^{d}$   \\
  &          & \multicolumn{2}{l}{202.03$^{k}$}&      &                                 	    	 & & 202.024$^{e}$   \\
  &          & 202.05\0&$\pm$  &\00.1$^h$               &                 			 & & 202.041$^{l}$ \\
  &          & 202.09\0&$\pm$  &\00.03$^g$            &                 			 & &              \\
 \\
 & $\Gamma$
             & 29.7\0\0&$\pm$  &\02.5$^{\dagger}$     & \030.53$^{a}$   & & \029.83$^{c}$    \\
 &           &         &       &                      			& \030.34$^{b}$   & & \029.80$^{d}$    \\
 &           &         &       &                  &                 	& 	& \030.52$^{e}$    \\
 &           &         &       &                  &                 	& 	& \030.60$^{l}$    \\
\\
 & $\overline{\sigma}^{\rm PI}$
             & 5.7\0\0 & $\pm$ & \01.2$^{\dagger}$    & \0\05.16$^{a}$  & &                  \\
 &           &         &       &                      & \0\05.37$^{b}$  & &                  \\
\br
\end{tabular}
~\\
$^{\dagger}$Present ALS experimental results\\
$^{a}$Breit-Pauli semi-relativistic intermediate coupling R-matrix (17-state).\\
$^{b}$Non-relativistic $LS$ coupling R-matrix (11-state).\\
$^{c}$Saddle-point method (SPM) with R-matrix \cite{wu91}.\\
$^{d}$Saddle-point method (SPM) with Complex Rotation \cite{gd00}\\
$^{e}$B-spline method \cite{Hansen2001}\\
$^{f}$Beam-foil spectroscopy and high-resolution spark spectroscopy \cite{Kramida2008} \\
$^g$Laser-produced plasmas experiment \cite{kennedy78}\\
$^h$Electron spectroscopy experimental data \cite{rodbro79} revised \cite{chung83}; the error bars of 0.1~eV are the estimates from the work of Chung and Bruch \cite{chung83} who provide the Auger energies with 5-digit numbers\\
$^{k}$Laser-produced plasmas experiment, no uncertainties specified \cite{Jannitti1984}\\
$^{l}$Saddle-point method (SPM) with relativistic corrections \cite{davis85}\\
\end{indented}
\end{table}

%
%
%

\begin{table}
    \caption{\label{tab:fit2} Resonance energies $E_{\rm ph}^{\rm (res)}$ (in eV),
             natural linewidths $\Gamma$ (meV) and resonance strengths $\overline{\sigma}^{\rm PI}$
            (in Mb~eV) for B$^{2+}$ ions from our present  experimental (ALS) and theoretical (R-matrix)
            work compared with previous studies for the higher lying resonances (peaks~\mycirc{3} through~\mycirc{7}) in the energy range 220--232~eV. For conversions from different energy scales of other original work to the present excitation energies see table~\ref{tab:fit1}.}
 \begin{indented}
 \lineup
 \item[]\begin{tabular}{@{}ccr@{\,}c@{\,}llcl}
 \br
 Resonance    & & \multicolumn{3}{c}{ALS/Others}   & \multicolumn{1}{c}{R-matrix} & \multicolumn{2}{c}{B-spline/SPM}\\
 (Label)      & & \multicolumn{3}{c}{(Experiment)} & \multicolumn{1}{c}{(Theory)} & \multicolumn{2}{c}{(Theory)}\\
 \ns
 \mr
 \lineup
  $$[(1s\,2s $^3$S)3p]$^2$P$\rm ^o$ & $E_{\rm ph}^{\rm (res)}$
           & 220.823 & $\pm$& 0.030$^{\dagger}$ & 220.721$^{a}$        & & 220.826$^b$ \\
  ~\mycirc{3}     &        & \multicolumn{2}{l}{220.73$^{c,*}$}      &&                      		 & &  220.851$^g$  \\
  &        & 220.766 & $\pm$& 0.074$^{d}$          &                      & &                \\
  &        & 220.77\0& $\pm$& 0.08$^{e,*}$         &                      & &                \\
  &        & 220.80\0& $\pm$& 0.2$^f$                 &                      & &                \\
  & $\Gamma$
           &         &       &                  	& \0\00.50$^{a}$     & & \0\00.46$^b$  \\
 &         &         &       &                     &      			     & & \0\00.38$^g$  \\

 & $\overline{\sigma}^{\rm PI}$
           & 6.5\0\0 & $\pm$ & 1.3$^{\dagger}$  & \0\08.99$^{a}$      & & \\
\\
 $$[(1s\,2s $^1$S)3p]$^2$P$\rm ^o$ & $E_{\rm ph}^{\rm (res)}$
           & 224.411 & $\pm$& 0.030$^{\dagger}$& 224.301$^{a}$      	 & & 224.416$^b$ \\
 ~\mycirc{4}     &         & \multicolumn{2}{l}{224.20$^c$}           & &                     	   	 & &  224.476$^g$ \\
 &         & 224.37\0& $\pm$& 0.08$^e$               &                    		 & &               \\
 &         & 224.374 & $\pm$& 0.037$^{d}$          &                    		 & &               \\
 &         & 224.41\0& $\pm$& 0.2$^f$                  &                     		 & &               \\
 & $\Gamma$
           &         &	     &                  & \0\00.28$^{a}$     & & \0\00.26$^b$ \\
 & $\overline{\sigma}^{\rm PI}$
           & 0.79\0& $\pm$ & 0.2$^{\dagger}$  & \0\01.88$^{a}$      & & \\
  \\
 $$[(1s\,2p $^3$P)3s]$^2$P$\rm ^o$ & $E_{\rm ph}^{\rm (res)}$
           &         &       &                & 225.407$^{a}$       				 & & 225.482$^b$  \\
  &        	& 225.416 & $\pm$& 0.005$^{d}$       &                     		& & 225.554$^g$\\
  &        & 225.44\0& $\pm$& 0.2$^f$        &                     & & \\
  & $\Gamma$
           &         &       &                  & \011.39$^{a}$      & & \011.39$^b$  \\
   &       &         &       &                     &      			     & & \011.32$^g$  \\
 \\
  $$[(1s\,2p $^1$P)3s]$^2$P$\rm ^o$ & $E_{\rm ph}^{\rm (res)}$
           & 226.834 & $\pm$& 0.030$^{\dagger}$ & 226.791$^{a}$       & & 226.849$^b$ \\
 ~\mycirc{5}      &        & 226.680 & $\pm$& 0.037$^{d}$       &                     & & \\
 & $\Gamma$
           & 	     &       &                  & \0\00.46$^{a}$      & & \0\00.47$^b$ \\
 & $\overline{\sigma}^{\rm PI}$
           & 1.5\0\0 &$\pm$& 0.3$^{\dagger}$   & \0\01.80$^{a}$      & & \\
 \\
  $$[(1s\,2p $^3$P)3d]$^2$P$\rm ^o$ & $E_{\rm ph}^{\rm (res)}$
           &         &      &			     	& 227.731$^{a}$       & & 227.760$^b$ \\
 & $\Gamma$
           &         &      &                   & \0\00.59$^{a}$     & & \0\00.46$^b$ \\
 \\
 $$[(1s\,2p $^1$P)3d]$^2$P$\rm ^o$ & $E_{\rm ph}^{\rm (res)}$
           & 228.394 & $\pm$ & 0.030$^{\dagger}$ & 228.371$^{a}$       & & \\
 ~\mycirc{6}     & $\Gamma$
           &         & 	     &                  & \0\00.66$^{a}$      & &	\\
 & $\overline{\sigma}^{\rm PI}$
           & 2.1\0\0 & $\pm$& 0.6$^{\dagger}$  & \0\02.03$^{a}$      & & \\
  \\
 $$[1s(3s\,3p) $^1$P]$^2$P$\rm ^o$ & $E_{\rm ph}^{\rm (res)}$
           & 231.233 & $\pm$& 0.030$^{\dagger}$& 231.201$^{a}$       & &  \\
  ~\mycirc{7}    & $\Gamma$
           &         &       &                  & \0\00.07$^{a}$     & &	\\
 & $\overline{\sigma}^{\rm PI}$
           & 0.8\0\0 & $\pm$& 0.3$^{\dagger}$  & \0\01.07$^{a}$      & & \\
\br
\end{tabular}
~\\
$^{\dagger}$Present ALS experimental results.\\
$^{a}$Breit-Pauli semi-relativistic intermediate coupling R-matrix (17-state).\\
$^b$B-spline method \cite{Hansen2001}\\
$^c$Laser-produced plasmas experiment, no uncertainties specified \cite{Jannitti1984}\\
$^{d}$Beam-foil spectroscopy and high-resolution spark spectroscopy \cite{Kramida2008} \\
$^e$Laser-produced plasmas experiment \cite{kennedy78}\\
$^f$Electron spectroscopy experimental data \cite{rodbro79} revised \cite{chung83}; the error bars of 0.2~eV are the estimates from the work of Chung and Bruch \cite{chung83} who provide the Auger energies with 5-digit numbers\\
$^g$Saddle-point method (SPM) with relativistic corrections \cite{davis85}\\
$^{*}$Line assignment changed as suggested by Kramida \etal \cite{Kramida2008}
\end{indented}
\end{table}

Table 1  displays the corresponding ALS experimental results together with
the present $LS$ and intermediate-coupling R-matrix calculations in addition to those
from the earlier experimental \cite{rodbro79,kennedy78,Jannitti1984,Kramida2008,chung83} and theoretical \cite{Hansen2001,wu91,gd00,davis85} studies.
Comparisons of the resonance parameters (see table 1)  for the [1s\,(2s\,2p)$^3$P]$^2$P$\rm ^o$ (peak~\mycirc{1})
and [1s\,(2s\,2p)$^1$P]$^2$P$\rm ^o$ (peak~\mycirc{2}) resonances determined experimentally with the current and previous theoretical estimates show a very satisfying level of agreement. The uncertainties of the experimental data for the excitation energies $E_{\rm ph}^{(res)}$ are limited by the existing calibration standards in the case of the present work and the previous study by Kennedy \etal~\cite{kennedy78} with error bars of 30~meV. The electron spectroscopy data have uncertainties of 100~meV, considerably higher than the photon experiments. The data provided by Kramida \etal~\cite{Kramida2008} have the character of recommended data and as such are based not only on existing experiments but also on theoretical calculations using {\sl ab initio} and semi-empirical techniques. The estimated uncertainties are only 10~meV and 14~meV for the energies of the [1s\,(2s\,2p)$^3$P]$^2$P$\rm ^o$ and [1s\,(2s\,2p)$^1$P]$^2$P$\rm ^o$ resonances, respectively. The present experiment is closest to this work for both peaks~\mycirc{1} and~\mycirc{2}.

All experimental $E_{\rm ph}^{(res)}$ results for peak~\mycirc{1} agree with one another within their uncertainties. The present R-matrix result is within about 50~meV from the most accurate experimental data. The theoretical data closest to experiment are those resulting from the configuration interaction expansion calculations using a B-spline basis set~\cite{Hansen2001}. The result of the saddle-point method with R-matrix calculations~\cite{wu91} has the largest discrepancy with respect to the photoexcitation experiments with a deviation of 130~meV from the present experimental result and 149~meV from the excitation energy recommended by Kramida \etal~\cite{Kramida2008}.   For peak~\mycirc{2} the experiment by Kennedy \etal~\cite{kennedy78} is outside the uncertainty ranges of the other photoexcitation experiments with a difference of about 100~meV to the present experimental results. The scatter of the theory data is similar to that observed for peak~\mycirc{1}. The agreement of the present $LSJ$ R-matrix calculation for peak~\mycirc{2} with the present experiment is perfect while the B-spline method is 37~meV above the present experiment and 48~meV above the energy recommended by Kramida \etal~\cite{Kramida2008}. It is interesting to note that the differences of resonance energies obtained for peaks~\mycirc{1} and~\mycirc{2} by theory and experiment show a considerably reduced scatter range, about one third,  as compared to the resonance energies themselves. Excellent agreement with respect to the energy splitting is found for the present experiment with 2.838$\pm$ 0.001~meV, the photoexcitation experiment by Jannitti \etal~\cite{Jannitti1984} (8~meV lower) and the data of Kramida \etal~\cite{Kramida2008} (8~meV higher). With respect to these experimental data the theoretical predictions on the average are 50~meV too high.

In the prediction of the Lorentzian width of peak~\mycirc{1} the different theoretical approaches, all within the span of different versions of the saddle-point method,  cover a range from 4.0~meV to almost 5.5~meV, i.e., they differ by up to about 30\% from the lowest width calculated. The present R-matrix calculations agree best with the experiment but also the saddle-point method with R-matrix \cite{wu91} calculations and the B-spline method \cite{Hansen2001} are within the present experimental error bars. For peak~\mycirc{2}  with its larger width all theoretical data and the present experiment are in very good accord. Since this is the first time that Lorentzian widths of core-excited  states of the B$^{2+}$ ion have been experimentally determined, there are no other measurements to compare with. In general, discrimination of specific theoretical approaches is not easily possible on the basis of the existing experimental data for $E_{\rm ph}^{(res)}$ and $\Gamma$ provided in table 1. In view of all available data, both theoretical and experimental, the excitation energies from the saddle-point method with R-matrix calculations~\cite{wu91} appear to be too high (at a level of 0.1~eV).

From the experimental resonance strengths also the oscillator strengths $f$ may be determined using the relationship,
\begin{equation}
f ( L_i S_i \rightarrow  L_j S_j ) = {\frac{\overline{\sigma}^{\rm PI}}{4 \pi^2 \alpha a_{0}^2 {\cal R}}}   \quad,
\end{equation}
with the fine structure constant $\alpha$, the first Bohr radius $a_0$ and the Rydberg energy ${\cal R}$.
For the two resonances in table 1,  the oscillator strengths obtained from
the experimental results yield values of 0.46 $\pm$  0.09  for the first resonance
at 199.149~eV and  0.052 $\pm$ 0.010 for the second resonance at 201.987~eV
which are in suitable agreement with our theoretical estimates of 0.41 ($LSJ$), 0.42 ($LS$) and 0.047 ($LSJ$), 0.049 ($LS$).
We note that any tiny contributions to the decay rate from radiative processes
are negligible compared to our absolute error bars. Thus the measured photoionization
resonances account for almost all of the oscillator strength in these transitions.

Table 2 gives the values determined for the resonance energies and resonance strength parameters
of the remaining higher lying resonances located in the energy region 220--238.3~eV along
with the corresponding theoretical values.  From table 2 (due to the limited scan ranges and the
resolution of the individual scans around 25~meV applied in the region 220--230~eV) it is seen that we were not able to
resolve and detect the two small resonances (i.e. the peak lying between \mycirc{4} and \mycirc{5}, and that between \mycirc{5} and ~\mycirc{6})
in the spectra for this energy range.  For the peaks that we have been able to observe experimentally
(and determine their resonance parameters and tentatively designate),  a comparison with theory
shows the results are in suitable agreement both with our theoretical estimates and those from earlier studies.

The Lorentzian widths of the resonances discussed in table~2 could not be experimentally determined. Most of the states have calculated widths between 0.07~meV and 0.66~meV, a range that is presently not experimentally accessible. Only the [(1s\,2p $^3$P)3s]$^2$P$\rm ^o$ resonance is predicted theoretically to have a width of 11.4 meV, sufficiently broad for the possible energy resolution of the experimental arrangement. However, this resonance is so small it could not even be detected in the measurements. Where a comparison is possible the Lorentzian widths resulting from the present R-matrix approach agree very well with the results of the B-spline method~\cite{Hansen2001} and saddle-point method with relativistic corrections~\cite{chung83}. Discrepancies are typically within less than 10\%.

The experimental resonance energies given in table 2 for peak~\mycirc{3} (assigned to the excited [(1s\,2s $^3$S)3p]$^2$P$\rm ^o$ state) are all within their mutual error bars. The scatter of the theoretical data is not more than 130~meV. For peak~\mycirc{4} (assigned to the excited [(1s\,2s $^1$S)3p]$^2$P$\rm ^o$ state) the experimental results apart from that by Jannitti\etal~\cite{Jannitti1984} (no error bars quoted, difference about 0.2~eV) agree very well with one another. The theory data are close to the experiments that quote the lowest uncertainties. For peaks \mycirc{3} through \mycirc{7}  the present experimental results for the resonance energies have the smallest uncertainty with a conservative estimate of $\pm$30~meV. For the [(1s\,2p $^3$P)3s]$^2$P$\rm ^o$ resonance, not seen in the present experiment, Kramida \etal~\cite{Kramida2008} estimated an uncertainty of only 5~meV of their resonance energy. This is barely in agreement with the present R-matrix results and would rule out the results from the B-spline method~\cite{Hansen2001} and saddle-point method with relativistic corrections~\cite{chung83} which differ by 66~meV and 139~meV, respectively.

For peak~\mycirc{5} (assigned to the excited [(1s\,2p $^1$P)3s]$^2$P$\rm ^o$ state) the present resonance energy and that of Kramida \etal~\cite{Kramida2008} differ by 154~meV, far outside the mutual error bars. On the other hand the available theoretical calculations agree with the present experiment within its error bar. Given the excellent agreement of the present energies and the data of Kramida \etal~\cite{Kramida2008} in all other cases one must assume a misprint in their table or a problem with the assignment of the state associated with the quoted energy. For the [(1s\,2p $^3$P)3d]$^2$P$\rm ^o$ resonance not seen in the present experiment, the present R-matrix resonance energy is within 31~meV from the result of the B-spline method~\cite{Hansen2001}. For the remaining peaks~\mycirc{6} and~\mycirc{7} (assigned to the excited [(1s\,2p $^1$P)3d]$^2$P$\rm ^o$ and [1s(3s\,3p) $^1$P]$^2$P$\rm ^o$ states, respectively) there are no data other than the present theory and experiment. They are compared with one another in the context of figures~4 and 5. Overall, the level-energy results of the B-spline method~\cite{Hansen2001}, where available, are found to be the  theoretical data closest to the most accurate experiments.

The theoretical data in figure 4 (for peaks~\mycirc{1} through~\mycirc{6}) was shifted by minor amounts
(details are given in the caption)  to match the experimental data.
In order to make a direct comparison with the absolute experimental data,
the theoretical cross sections were convoluted with  Gaussians of the appropriate widths.
In figure 5 we show the comparison between theory and experiment
in the energy range 230.0--238.3~eV, where the theoretical data  has been convoluted
with a Gaussian of 135~meV FWHM.  Finally, figure 6 illustrates the comparison between the
present R-matrix  intermediate coupling calculations and the ALS experimental results over
the energy range incorporating the 7 peaks observed in the experiment.
The spectral overview in figure 6 illustrates a simulated ALS spectrum expected on the basis
of the resonance parameters from the present measurements and their analysis assuming
the cross sections were all measured with 135~meV resolution
(the experimental spectrum displayed in figure 5 was indeed measured at 135~meV resolution).
As can be seen from figure 6 over the complete energy region studied excellent agreement  is observed
between the present theoretical R-matrix calculations performed in intermediate coupling
and the experimental data taken at the ALS. The figure also indicates the resonances that were not observed in the experiments because of their small cross sections.

\section{Conclusion}
Photoionization of B$^{2+}$ ions was studied both experimentally and theoretically in the energy  region of the K-edge.
Overall, excellent agreement is found between the present theoretical and
experimental results both on the photon-energy scale and on the
absolute photoionization cross-section scale for this prototype Li-like system for the majority of the observed peaks.
Minor discrepancies however do exist between theory and experiment
(when the energy range around narrow resonances is strongly expanded) both for resonance strengths and positions.
This may be possibly attributed to the limitations of the n=4 basis set used in the present theoretical work.

The strength of the present study is in its excellent
experimental resolving power coupled
with theoretical predictions using the Breit-Pauli R-matrix method.
The experimental energy resolution of 11.3~meV and 14.3~meV for the
first two peaks in the present work made possible the determination of the
linewidths of the [1s\,(2s\,2p)$^{1,3}$P]$^2$P$\rm ^o$ resonances.
The Voigt line-profile fit (to the ALS experimental data)
 for these two resonances yielded values respectively  (see table \ref{tab:fit1}) for the
linewidth of $4.8 \pm 0.6$~meV  and $29.7 \pm 2.5$~meV
which are in good agreement with the present R-matrix
theoretical predictions of  4.40~meV  and 30.53~meV
(5.27~meV and 29.53~meV from the saddle-point-method  with R-matrix \cite{wu91}
 and 4.0~meV and 29.8~meV from the saddle-point-method with complex
 co-ordinate rotation \cite{gd00}). Our linewidth results are also in excellent
 accord with the values of 4.29~meV and 30.52~meV  from the B-spline
 method \cite{Hansen2001} for these same two resonances.

\ack
We acknowledge support by
Deutsche Forschungsgemeinschaft under project number Mu 1068/10  and through
NATO Collaborative Linkage grant 976362 as well as by the US Department of Energy (DOE)
under contract DE-AC03-76SF-00098 and grant  DE-FG02-03ER15424.
B M McLaughlin acknowledges support by the US
National Science Foundation through a grant to ITAMP
at the Harvard-Smithsonian Center for Astrophysics.
The computational work was carried out at the National Energy Research Scientific
Computing Center in Oakland, California USA and on the Tera-grid at
the National Institute for Computational Science (NICS) in Tennessee USA,
which is supported in part by the US National Science Foundation.


%
%
%
%
~\\~\\

\bibliographystyle{iopart-num}

\bibliography{b2plus}

\providecommand{\newblock}{}
\begin{thebibliography}{10}
\expandafter\ifx\csname url\endcsname\relax
  \def\url#1{{\tt #1}}\fi
\expandafter\ifx\csname urlprefix\endcsname\relax\def\urlprefix{URL }\fi
\providecommand{\eprint}[2][]{\url{#2}}

\bibitem{McLaughlin2001}
{McLaughlin B M} 2001 {} {\em {Spectroscopic Challenges of Photoionized
  Plasma}\/} ({\em ASP Con$f$. Series\/} vol \textbf{247}) ed {Ferland, G and
  Savin D W} (San Francisco, CA: Astronomical Society of the Pacific) p~87

\bibitem{bruch77}
{B}ruch R, {R}{\o}dbro M, {B}isgaard P and {D}ahl P 1977 {\em {Phys. Rev.
  Lett.}\/} {\bf \textbf{39}} 801

\bibitem{rodbro77}
{R}{\o}dbro M, {B}ruch R, {B}isgaard P, {D}ahl P and {F}astrup B 1977 {\em {J.
  Phys. B: At. Mol. Phys.}\/} {\bf \textbf{10}} L483

\bibitem{rodbro79}
{R}{\o}dbro M, {B}ruch R and {B}isgaard P 1979 {\em {J. Phys. B: At. Mol.
  Phys.}\/} {\bf \textbf{12}} 2413

\bibitem{kennedy78}
{Kennedy E T and Carroll P K} 1978 {\em {J. Phys. B: At. Mol. Phys.}\/} {\bf
  \textbf{11}} 965

\bibitem{Jannitti1984}
Jannitti E, Nicolosi P and Tondello G 1984 {\em Physica C\/} {\bf 124} 139

\bibitem{Hofmann1990}
{Hofmann G, M\"{u}ller A, Tinschert K and Salzborn E} 1990 {\em {Z. Phys. D:
  Atoms, Molecules and Clusters}\/} {\bf \textbf{16}} 113

\bibitem{Kramida2008}
{Kramida A E, Ryabtsev A N, Ekberg J O, Jink I, Mannervik S and Martinson I}
  2008 {\em {Phys. Scr.}\/} {\bf \textbf{78}} 025301

\bibitem{safronova78}
{Vainshtein~L~A and Safronova~U~I } 1978 {\em {At. Data Nucl. Data Tables}\/}
  {\bf \textbf{21}} 49

\bibitem{Hansen2001}
{Verbockhaven G and Hansen J E} 2001 {\em {J. Phys. B: At. Mol. Opt. Phys.}\/}
  {\bf 34} 2337

\bibitem{chung83}
{Chung C T and Bruch~R} 1983 {\em {Phys. Rev. A}\/} {\bf \textbf{28}} 1418

\bibitem{wu91}
{Wu L and Xi J} 1991 {\em {J. Phys. B: At. Mol. Phys.}\/} {\bf \textbf{24}}
  3351

\bibitem{gd00}
{Bingcong G and Wensheng D} 2000 {\em {Phys. Rev. A}\/} {\bf \textbf{62}}
  032705

\bibitem{Fuhr2010}
{Fuhr J and Wiese W L} 2010 {\em {J. Phys. Chem. Ref. Data}\/} {\bf
  \textbf{39}} 013101

\bibitem{Schlachter2004}
{Schlachter A S, Sant'Anna M M, Covington A M, Aguilar A, Gharaibeh M~F, Emmons
  E D, Scully S W J, Phaneuf R A, Hinojosa G, {\'A}lvarez I, Cisneros C,
  M\"{u}ller A and McLaughlin B M} 2004 {\em {J. Phys. B: At. Mol. Opt.
  Phys.}\/} {\bf \textbf{37}} L103

\bibitem{Scully2005}
{Scully S W J, Aguilar A, Emmons E D, Phaneuf R A, Halka M, Leitner D, Levin J
  C, Lubell M S, P\"{u}ttner R, Schlachter A S, Covington A M, Schippers S,
  M\"{u}ller A and McLaughlin B M} 2005 {\em {J. Phys. B: At. Mol. Opt.
  Phys.}\/} {\bf \textbf{38}} 1967

\bibitem{Mueller2009}
{M{\"u}ller A, Schippers S, Phaneuf R~A, Scully S W J, Aguilar A, Covington A
  M, {\'A}lvarez I, Cisneros C, Emmons E D, Gharaibeh M~F, Schlachter A~S,
  Hinojosa G, and McLaughlin B~M} 2009 {\em {J. Phys. B: At. Mol. Opt.
  Phys.}\/} {\bf 42} 235602

\bibitem{Kasstra1993}
{Kaastra J S and Mewe R} 1993 {\em {Astron. \& Astrophys. Suppl. Ser.}\/} {\bf
  \textbf{97}} 443

\bibitem{rm1979}
{Reilman R and Manson S T} 1979 {\em {Astrophys. J. Suppl. Ser.}\/} {\bf
  \textbf{40}} 815

\bibitem{verner1993}
{Verner D A, Yakovlev D G, Band I M and Trzhaskovskaya M B} 1993 {\em {At. Data
  Nucl. Data Tables}\/} {\bf \textbf{55}} 233

\bibitem{verner1995}
{Verner D A and Yakovlev D G} 1995 {\em {Astron. \& Astrophys. Suppl. Ser.}\/}
  {\bf \textbf{109}} 125

\bibitem{Ferland1998}
{Ferland G J, Korista K T, Verner D A, Ferguson J W, Kingdon J B and Verner E
  M} 1998 {\em {Pub. Astron. Soc. Pac.(PASP)}\/} {\bf \textbf{110}} 761

\bibitem{Ferland2003}
{Ferland G J} 2003 {\em {Ann. Rev. of Astron. \& Astrophys.}\/} {\bf
  \textbf{41}} 517

\bibitem{Kallman2001}
{Kallman T R and Bautista M A} 2001 {\em {Astrophys. J. Suppl. Ser.}\/} {\bf
  \textbf{134}} 139

\bibitem{Cunha97}
{Cunha K, Lambert D L, Lemke M, Gies D R, Roberts L C} 1997 {\em {Astrophys.
  J.}\/} {\bf \textbf{478}} 211

\bibitem{Proffitt99}
{Proffitt C R, J{\"o}nsson P, Litzi{\'e}n U, Pickering J C and Wahlgren G M}
  1999 {\em {Astrophys. J.}\/} {\bf \textbf{516}} 342

\bibitem{Cunto1993}
{Cunto W, Mendoza C, Ochsenbein F and Zeippen C J} 1993 {\em {Astron. \&
  Astrophys.}\/} {\bf \textbf{275}} L5

\bibitem{safronova69}
{Safronova~U~I and Kharitonova~V~N} 1969 {\em {Opt. Spectrosc}\/} {\bf
  \textbf{27}} 300

\bibitem{goldsmith74}
{Goldsmith~S} 1974 {\em {J. Phys. B: At. Mol. Phys.}\/} {\bf \textbf{7}} 2315

\bibitem{bhatia78}
{Bhatia A K} 1978 {\em {Phys. Rev. A}\/} {\bf \textbf{18}} 2523

\bibitem{davis85}
{Davis B F and Chung~K~T} 1985 {\em {Phys. Rev. A}\/} {\bf \textbf{31}} 3017

\bibitem{Scully2006}
{Scully S W J, {\'A}lvarez I, Cisneros C, Emmons E D, Gharaibeh M~F, Leitner D,
  Lubell M S, M\"{u}ller A, Phaneuf R A, P\"{u}ttner R, Schlachter A S,
  Schippers S, Ballance C P and McLaughlin B M} 2006 {\em {J. Phys. B: At. Mol.
  Opt. Phys.}\/} {\bf 39} 3957

\bibitem{Scully2007}
{Scully S W J, {\'A}lvarez I, Cisneros C, Emmons E D, Gharaibeh M~F, Leitner D,
  Lubell M S, M\"{u}ller A, Phaneuf R A, P\"{u}ttner R, Schlachter A S,
  Schippers S, Ballance C P and McLaughlin B M} 2007 {\em {J. Phys. Conf. Ser.
  }\/} {\bf \textbf{58}} 387

\bibitem{damp}
{Robicheaux F, Gorczyca T W, Griffin D C, Pindzola M S and Badnell N R} 1995
  {\em {Phys. Rev. A}\/} {\bf 52} 1319

\bibitem{rmat}
{Burke P G and Berrington K A} 1993 {\em {Atomic and Molecular Processes: An
  R-matrix Approach}\/} (Bristol, UK: IOP Publishing)

\bibitem{codes}
{Berrington K A, Eissner W and Norrington P~H} 1995 {\em {Comput. Phys.
  Commun.}\/} {\bf \textbf{92}} 290
  \urlprefix\url{http://amdpp.phys.strath.ac.uk/APAP}

\bibitem{Hibbert1975}
{Hibbert A} 1975 {\em {Comput. Phys. Commun.}\/} {\bf \textbf{9}} 141

\bibitem{Clementi1974}
{Clementi E and Roetti C} 1974 {\em {At. Data Nucl. Data Tables}\/} {\bf
  \textbf{14}} 177

\bibitem{Ralchenko2008}
{Ralchenko Y, Kramida A E, Reader J, and NIST ASD Team,} {NIST Atomic Spectra
  Database (version 3.1.4),} National Institute of Standards and Technology,
  Gaithersburg, MD, USA \urlprefix\url{http://physics.nist.gov/asd3}

\bibitem{keith1996}
{Quigley L and Berrington K~A} 1996 {\em {J. Phys. B: At. Mol. Phys.}\/} {\bf
  \textbf{29}} 4529

\bibitem{keith1998}
{Quigley L, Berrington K A and Pelan J} 1998 {\em {Comput. Phys. Commun.}\/}
  {\bf 114} 225

\bibitem{keith1999}
{Ballance C P, Berrington K A and McLaughlin B M} 1999 {\em {Phys. Rev. A}\/}
  {\bf \textbf{60}} R4217

\bibitem{Covington2002}
{Covington A~M, Aguilar A, Covington I~R, Gharaibeh M~F, Shirley C~A, Phaneuf
  R~A, {\'A}lvarez I, Cisneros C, Hinojosa G, Dominguez-Lopez I, Sant'Anna M~M,
  Schlachter A~S, McLaughlin B~M and Dalgarno} 2002 {\em {Phys. Rev. A}\/} {\bf
  \textbf{66}} 062710

\bibitem{Davis1989}
Davis B~F and Chung K~T 1989 {\em Phys. Rev. A\/} {\bf 39} 3942

\bibitem{Broetz2001}
{Br{\"o}tz F, Trassl R, McCullough R~W, Arnold W and Salzborn E} 2001 {\em
  {Phys. Scr.}\/} {\bf \textbf{T92}} 278

\bibitem{Hudson1993}
Hudson E, Shirley D~A, Domke M, Remmers G, Puschmann A, Mandel T, Xue C and
  Kaindl G 1993 {\em {Phys. Rev. A}\/} {\bf 47} 361

\bibitem{King1977}
{King G C, Tronc M, Read F H, and Bradford R C} 1977 {\em {J. Phys. B: At. Mol.
  Phys.}\/} {\bf \textbf{10}} 2479

\end{thebibliography}

\end{document}